# Chemical Vapor Deposition Growth of Monolayer WSe$_2$ with Tunable Device Characteristics and Growth Mechanism Study

Bilu Liu[+]*, Mohammad Fathi[+], Liang Chen, Ahmad Abbas, Yuqiang Ma, Chongwu Zhou*

Ming Hsieh Department of Electrical Engineering, University of Southern California, Los Angeles, California 90089, USA

[+]Equal contribution

Address correspondence to: chongwuz@usc.edu, biluliu@usc.edu

**Abstract**

Semiconducting transition metal dichalcogenides (TMDCs) have attracted a lot of attention recently, because of their interesting electronic, optical, and mechanical properties. Among large numbers of TMDCs, monolayer of tungsten diselenides (WSe$_2$) is of particular interest since it possesses a direct band gap and tunable charge transport behaviors, which make it suitable for a variety of electronic and optoelectronic applications. Direct synthesis of large domains of monolayer WSe$_2$ and their growth mechanism studies are important steps toward applications of WSe$_2$. Here, we report systematical studies on ambient pressure chemical vapor deposition (CVD) growth of monolayer and few layer WSe$_2$ flakes directly on silica substrates. The WSe$_2$ flakes were characterized using optical microscopy, atomic force microscopy, Raman spectroscopy, and photoluminescence spectroscopy. We investigated how growth parameters, with emphases on growth temperatures and durations, affect the sizes, layer numbers, and shapes of as-grown $WSe_2$ flakes. We also demonstrated that transport properties of CVD-grown monolayer $WSe_2$, similar to mechanically-exfoliated samples, can be tuned into either p-type or ambipolar electrical behavior, depending on the types of metal contacts. These results deepen our understandings on the vapor phase growth mechanism of $WSe_2$, and may benefit the uses of these CVD-grown monolayer materials in electronic and optoelectronics.

Two-dimensional (2D) layered atomic crystals have drawn significant interest in the past decade. Semi-metallic graphene with a zero electronic band gap and insulating hexagonal boron nitride are two important materials in this family that have been extensively studied. There is growing interest in recent years to explore 2D materials possess semiconducting properties for potential usage in electronics and optoelectronics.[1-3] Transitional metal dichalcogenides (TMDCs) are a large family of materials with tunable electronic properties. For example, band gaps of semiconducting TMDCs can be tuned *via* tuning layer numbers, chemical compositions, strains of the materials, *etc*.[1, 2, 4, 5] Significant research efforts have been devoted towards synthesis, structure and defect characterization, and electronic and optoelectronic applications of TMDCs, especially semiconducting TMDCs.[5-18]

Among various semiconducting TMDCs, $MoS_2$ is the one which attracts most attention. Growth of large crystals, continuous films, and patterned growth of $MoS_2$ monolayers and few layers have been reported in the past few years.[6-9, 11, 12, 17, 19-23] Moreover, field effect transistors (FETs) fabricated using both mechanically-exfoliated and vapor phase grown $MoS_2$ have been widely studied. Typically, $MoS_2$ transistors show a n-type behavior,[10, 24-26] and in some special cases, p-type $MoS_2$ transistors have also been demonstrated, for example, by using high work function $MoO_{3-x}$ as source and drain contacts.[27]

$WSe_2$ is another interesting TMDC material as it exhibits unique and complementary properties to the prototype TMDC material, $MoS_2$. Monolayer $WSe_2$ has a band gap smaller than monolayer $MoS_2$ (~1.65 eV for monolayer $WSe_2$ versus 1.8 eV for monolayer $MoS_2$), while they possess similar band gaps in bulk form (1.2 eV for both). Importantly, it has been demonstrated that transport properties of mechanically-exfoliated monolayer $WSe_2$ can be facilely tuned to be either p-type or ambipolar behavior, depending on the types of contact metals.[28] $WSe_2$ is also a material that has a high absorption coefficient in the visible to infrared range, a high quantum yield in photoluminescence (PL), and a strong spin-orbit coupling.[1, 29, 30] Taking these advantages, a variety of optoelectronic devices such as photodetectors, light-emitting diodes, and photovoltaic devices have been made using mechanically-exfoliated monolayer $WSe_2$ by using split gate device structures.[30-32] Another interesting property of $WSe_2$ is that this material holds the lowest thermal conductivity among dense solid in disordered films of layered WSe$_2$ crystals, which may find applications as thermoelectric materials.[33] Property controlled synthesis of WSe$_2$ is a prerequisite for its applications in many fields. There have been a few recent reports toward vapor phase growth of thin WSe$_2$ flakes.[29, 34-37] For example, Li *et al.* reported growth of large area monolayer WSe$_2$ on sapphire substrates by low-pressure chemical vapor deposition (CVD).[35] Top gated FETs fabricated using these WSe$_2$ samples showed ambipolar transport behavior. More recently, direct vapor phase sublimation of WSe$_2$ powders has been demonstrated to be able to produce WSe$_2$ monolayers. Xu *et al.* show that such vapor phase grown WSe$_2$ shows similar optical quality with exfoliated samples.[29] In another study, Duan *et al.* showed that back gated FETs fabricated using vapor phase grown WSe$_2$ exhibiting unipolar p-type behavior,[34] similar to a study by Xiang *et al.*,[37] and is different with ambipolar behavior shown in CVD-grown WSe$_2$ monolayer samples.[35] Growth of WSe$_2$ by metal organic CVD method has also been reported recently, yielding WSe$_2$ monolayers with domain sizes ranging from a few hundreds of nanometers to the largest size of 8 μm.[38] Nevertheless, even with the above achievements, the growth of WSe$_2$ is still much less studied than other TMDCs like MoS$_2$, and growth mechanisms of WSe$_2$ remain poorly understood. It is known that WSe$_2$ is relatively difficult to synthesize than MoS$_2$, due to the fact that selenium precursors are less reactive than sulfur precursors. In addition, metal oxides are typical source materials for CVD growth of TMDCs. In the cases of MoS$_2$ and WSe$_2$, WO$_3$ is much more difficult to sublimate than MoO$_3$, due to their large difference in boiling points and consequently, vapor pressures (the boiling points of WO$_3$ and MoO$_3$ are 1700 °C and 1155 °C, respectively).[39] Currently, how the influence of various growth parameters on the properties of as-grown WSe$_2$ flakes, including layer numbers, shapes, and sizes, remain poorly understood. In addition, vapor phase grown WSe$_2$ usually exhibits either p-type or ambipolar transport behavior in different reports,[34, 35] while the tunability of such transport behavior in the same material has rarely been reported so far. Therefore, it is fundamentally important to know whether vapor phase synthesized WSe$_2$ could deliver similar tunability in charge transport properties compared to mechanically-exfoliated samples. In this contribution, we performed systematical experiments to study CVD growth of monolayer and few layer WSe$_2$ and how various growth parameters, especially growth temperatures and growth durations, affect the properties of as-grown WSe$_2$ flakes in terms of their sizes, shapes, and thicknesses. Moreover, transport studies showed that the vapor phase grown WSe$_2$ monolayer can exhibit either p-type or ambipolar transport behavior, depending on the types of metal contacts used, suggesting the use of these CVD-grown samples for electronics and optoelectronics.

**Results and Discussion**

The details of our CVD process are described in the Methods. Briefly, selenium (Se) powders and WO$_3$ powders were used as precursors for Se and W, respectively. The temperatures and distances of these two sources were carefully controlled and adjusted. The growth substrates were silicon with 300 nm thermally-grown $SiO_2$, and were placed at the position of $WO_3$ powders and facing down. In our experiments, several important CVD parameters were systematically studied and optimized. Among these parameters, we found that growth temperatures and durations are the two that have very significant influence on $WSe_2$ growth. In this regard, the growth temperatures (the temperature of $WO_3$ source and growth substrates) were tuned in a wide range between 800 and 1100 °C and their effects on the properties of as-grown $WSe_2$ were carefully studied. The effect of growth durations on the size and morphology of $WSe_2$ were also examined. The underlying mechanisms of how these parameters affect $WSe_2$ growth were discussed.

Figure 1 shows overall characterization of monolayer $WSe_2$ grown from optimized condition. As can be seen in Figure 1a, the dark colored flakes exhibit uniform triangular shape with edge lengths in the range of 5-20 μm. The triangular shape of $WSe_2$, $MoS_2$, and some other TMDCs is related to their crystal symmetry properties. Figure 1b and 1c are typical atomic force microscopy (AFM) image and a height profile of a thin flake, which shows a step height of ~0.9 nm. This height value is typical for monolayer TMDCs which is comprised of three atomic layers of X-M-X, as has been widely reported in both mechanically-exfoliated and CVD-grown monolayer TMDCs like $MoS_2$, $WSe_2$, etc.[6, 8, 28]

As fast and non-damaging techniques, Raman and PL spectroscopies are important tools to reveal layer numbers, optical quality, and strain information of $WSe_2$.[40] Figure 1d presents three typical Raman spectra collected from the same $WSe_2$ flake at different positions, as indicated in the optical image shown in the inset. The absence of the $B_{2g}^1$ peak at ~304 cm$^{-1}$ suggests that the flakes grown at this condition are monolayers, consistent with AFM results shown in Figure 1b and 1c. Figure 1d also reveals that the Raman spectra of $WSe_2$ flakes are uniform as the three spectra are nearly identical to each other, without any detectable difference in peak frequencies and negligible variations in peak intensities and full-width at half maximum (FWHM) of each peak. Figure 1f is a typical Raman intensity map of a monolayer $WSe_2$ flake. It further shows a quite uniform intensity over the whole flake. We have taken Raman mapping for more than ten $WSe_2$ flakes and all of them show uniform intensity over the whole flakes. Figure 1e and 1g are typical PL spectra and PL mapping of monolayer $WSe_2$. From Figure 1e, it can be seen that the PL peak positions (FWHM) of the black, blue, and red spectra are 1.58 eV (0.08 eV), 1.59 eV (0.06 eV), and 1.62 eV (0.07 eV), respectively. Bright light emission at ~1.60 eV and symmetric single PL peak suggest the direct band gap nature of monolayer $WSe_2$, showing good agreement with other recent reports about PL of monolayer $WSe_2$.[1, 29-32, 34, 35] We found that PL is a more sensitive tool than Raman to reveal the fine structure information of as-grown samples, because certain difference in PL intensity and peak positions can be found at different sample locations, as shown in Figure 1e and PL intensity map in Figure 1g. Similar non-uniform PL intensity map of vapor phase grown WSe$_2$ can also be seen in some recent papers.[5, 21, 29] There are several underlying mechanisms which may lead to the observed non-uniformity in PL, including strain, structural defects, doping, surface absorbates, *etc*.[41] Since Raman spectra of WSe$_2$ is sensitive to the doping and strain of the materials,[4] uniform Raman intensity map suggests that doping and strain should not be the key reason for the observed non-uniform PL of our WSe$_2$. Although it has not been well studied in WSe$_2$ case, recent results on MoS$_2$ do show that sulfur vacancy is one of the major defects in MoS$_2$ because of their low formation energy.[42-44] We performed energy dispersive spectroscopy analysis of as-grown WSe$_2$ flakes, and found that Se/W atomic ratio is less than 2. This result suggests an insufficiency of Se in the sample, which may be related to the formation of Se vacancies, similar to the defects in MoS$_2$. In collection with other recent studies about structural defects of CVD-grown TMDC materials and the origin of non-uniform PL from CVD-grown MoS$_2$ and WS$_2$,[11, 29, 41, 45] we speculate that structural defects, *e.g.*, Se vacancy, might be the reason for the non-uniform PL spectra observed here.

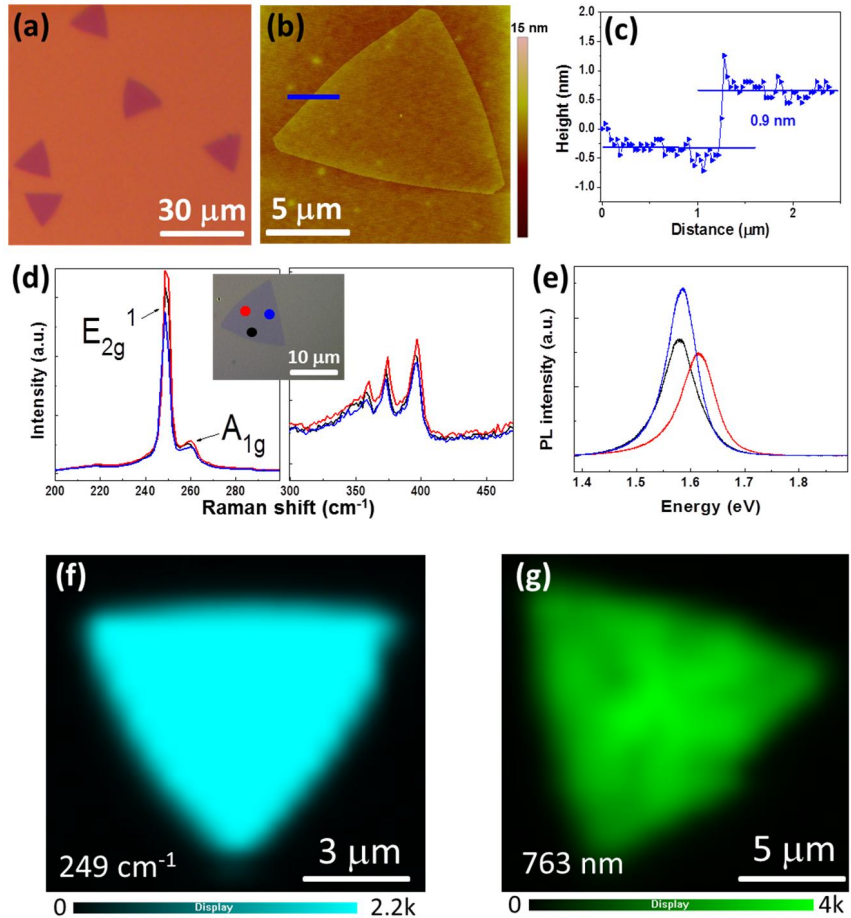

**Figure 1. CVD growth and microscopic characterization of monolayer WSe$_2$. (a) Optical microscopy image, (b) AFM image, and (c) Height profile of CVD-grown monolayer WSe$_2$ along the blue line in (b). (d) Raman and (e) PL spectra collected from the same monolayer WSe$_2$ flake at three different positions (indicated by different colors in the inset of Figure 1d). (f) Representative Raman and (g) PL intensity maps of monolayer WSe$_2$ flakes.**

Compared with MoS$_2$, one interesting feature of WSe$_2$ is its tunable transport behavior. Javey *et al.* have shown that mechanically-exfoliated monolayer WSe$_2$ exhibits either p-type or ambipolar behavior, depending on the types of contact metals.[28] Different behaviors were shown for vapor phase grown monolayer $WSe_2$. For example, Li *et al.* grew $WSe_2$ monolayers by CVD on sapphire, and their devices showed ambipolar behavior when using ionic gels as dielectric and Ni/Au as contact, in a top gate configuration.[35] In another study, Duan *et al.* grew $WSe_2$ monolayers by direct sublimation of $WSe_2$ powders and their devices exhibited unipolar p-type behavior when using Au as contacts in a back gate configuration.[34] We characterized charge transport properties of our CVD-grown $WSe_2$, and showed that their transport properties can be tuned by changing contact metals. We tried three different kinds of metal contacts, including Pd/Ti (50 nm Pd with 0.5 or 1 nm Ti underneath as adhesion layer), Au/Ti (50 nm Au with 0.5 or 1 nm Ti adhesion layer), and Ti/Au (5 nm Ti with 50 nm Au on top). For Pd/Ti or Au/Ti electrodes, since the thickness of Ti layers are only 0.5 nm to 1 nm, and such thicknesses are not sufficient to form a continuous film. We expect that the contact property was mainly determined by Pd or Au, not by the adhesion Ti layer. Schematic of a back gated CVD $WSe_2$ device is shown in Figure 2a and an optical microscopy image of a $WSe_2$ transistor is shown in Figure 2b. Figure 2c shows the transfer curves ($I_{ds}$-$V_{gs}$) of a typical Pd/Ti (50 nm/1 nm) contacted device, which exhibits unipolar p-type behavior, as can be seen from the semi-log scale plot (black curve in Figure 2c). As a comparison, Figure 2d is the transfer curves of a typical Au/Ti (50 nm/1 nm) contacted device, which exhibits ambipolar behavior with comparable conductance at p and n-branches. The devices contacted using Ti/Au (5 nm/50 nm) show ambipolar behavior, similar to Au/Ti-contacted devices as shown in Figure 2d. These results reveal that similar to mechanically-exfoliated monolayer WSe$_2$,[28] CVD grown WSe$_2$ can also deliver different transport behaviors which can be tuned by the types and properties of metal electrodes. These results show the potential for using CVD-grown large area monolayer WSe$_2$ for optoelectronic applications where ambipolar transport behavior is appreciated.[30-32]

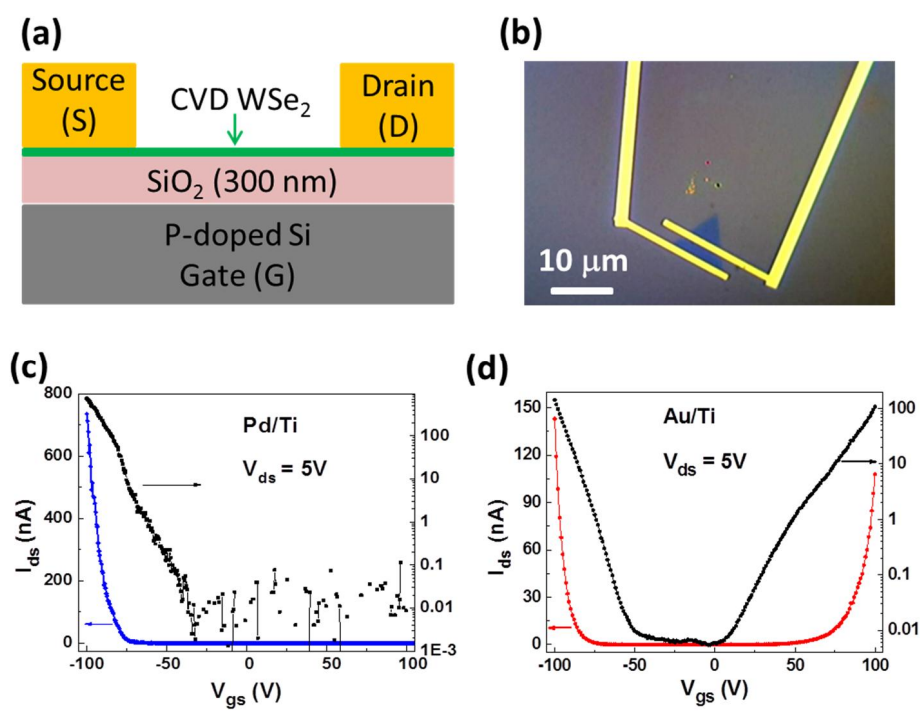

**Figure 2. Tunable device characteristics of CVD-grown monolayer WSe$_2$. (a) Schematic of a back gated WSe$_2$ transistor fabricated on Si/SiO$_2$ substrates where WSe$_2$ monolayers grown on. (b) An optical image of a monolayer WSe$_2$ transistor. (c) Representative transfer characteristics (I$_{ds}$-V$_{gs}$) of WSe$_2$ transistors using Pd/Ti (50 nm/1 nm) as source/drain metal contacts. (d) Representative transfer characteristics of WSe$_2$ devices using Au/Ti (50 nm/1nm) as source/drain metal contacts. The devices with Ti/Au (5 nm/50 nm) contacts exhibit similar behaviors with curves shown in (d).**

In our experiments, we found that many growth parameters have influence on the growth behavior of WSe$_2$, including yield, flake size, number of layers, and shapes of as-grown WSe$_2$. After a systematical exploration of a broad parameter space, we found that among all parameters, growth temperature is the most significant one. We first studied how growth temperature affects the properties of WSe$_2$ flakes.

Figure 3a-3c shows optical microscopy images of WSe$_2$ flakes grown at 850, 900, and 1050 °C, respectively. We observed several trends as the growth temperature increases. First, the flake sizes increase with increasing growth temperatures. A summary of average flake sizes at different growth temperatures is shown in Figure 3d, based on the analysis of several hundreds of flakes. Second, as the growth temperature increases, the layer numbers of WSe$_2$ also increase. Specifically, at growth temperatures below 950 °C, the as-grown WSe$_2$ flakes are nearly exclusively monolayers. However, when the growth temperatures are lower than 850 °C, the products are mainly particles together with a few numbers of very small flakes. Raman studies show that these particles are WSe$_2$. On the other hand, at growth temperatures above or equal to 1050 °C, all of the flakes observed were few layers. Some bilayer samples were found at temperatures between 950 and 1050 °C. Third, we found that the growth temperatures also have a critical influence on the shapes of as-grown WSe$_2$ flakes (Figure 4). In general, these experimental observations show

good agreement with a recent study on the growth of WSe$_2$ flakes by sublimation of WSe$_2$ powders.[34]

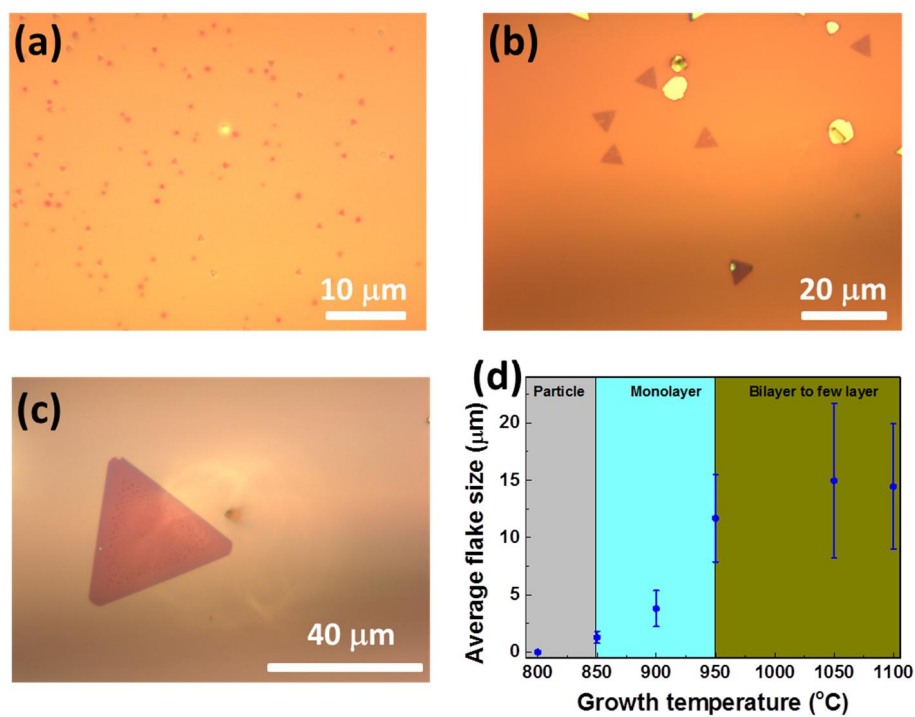

**Figure 3. Effect of growth temperatures on the sizes and layer numbers of CVD-grown WSe$_2$. Optical microscopy images of WSe$_2$ flakes grown at (a) 850 °C, (b) 900 °C, and (c) 1050 °C. The growth durations are 15 min for all cases. (d) The correlation of average WSe$_2$ flake sizes and layer numbers with growth temperatures. The vertical error bars indicate standard deviations of the flake sizes in statistical analysis.**

Shape is an important merit for 2D materials as it is related to their detailed atomic edge structures, which is determined by edge energetics at specific conditions. The effects of edge structures on the electronic, magnetic, and catalytic properties of 2D materials including graphene and TMDCs are expected.[46] Here we used AFM to study the shapes of as-grown WSe$_2$ flakes at different temperatures and to understand how the shape evolves with changing growth temperatures. Figure 4 presents amplitude AFM images of WSe$_2$ flakes grown at different temperatures ranging from 900 to 1050 °C. For monolayer flakes (growth temperatures of 850, 900, and 950 °C), we always observe that the samples are regular triangles (Figure 4a, 4b). When the temperatures increase to 1025 °C and beyond, some interesting features appear besides regular triangles. Figure 4c shows a thin few layer WSe$_2$ flake with a truncated triangular shape grown at 1025 °C. The white dotted lines in Figure 4c indicate the projected shape of the triangle. Interestingly, one can see that the short edges of the truncated triangle are not straight; instead, they are curved edges. This observation is different from a very recent study on the growth of truncated triangular MoS$_2$ flakes.[45] Such curved edges indicate that the atoms at each curved edge may not be made of the same element. This is because for edges solely terminated by either W or Se, they should be straight lines. Growth of WSe$_2$ flakes with different edge terminations provide material basis for edge-related property studies, such as their catalytic activity. Hexagonal WSe$_2$ flakes can also be found at the growth temperature of 1025 °C, as shown in Figure 4d. Again, with curved edges. When we further increase the growth temperatures to 1050 °C, even thicker flakes were grown. Figure 4e shows an AFM image of a truncated triangle shaped WSe$_2$ few layer. In this case, the short edges are quite straight, which is different with Figure 4c and 4d as in thin few layer samples. An AFM image of a nearly perfect hexagonal few layer WSe$_2$ flake with straight edges is also exhibited in Figure 4f. Such a hexagonal shape suggests that the adjacent edges are made of different atoms. Raman and PL measurements of the same flake were shown in Figure S1 in the Supporting Information, confirming it is a few layer WSe$_2$ flake.

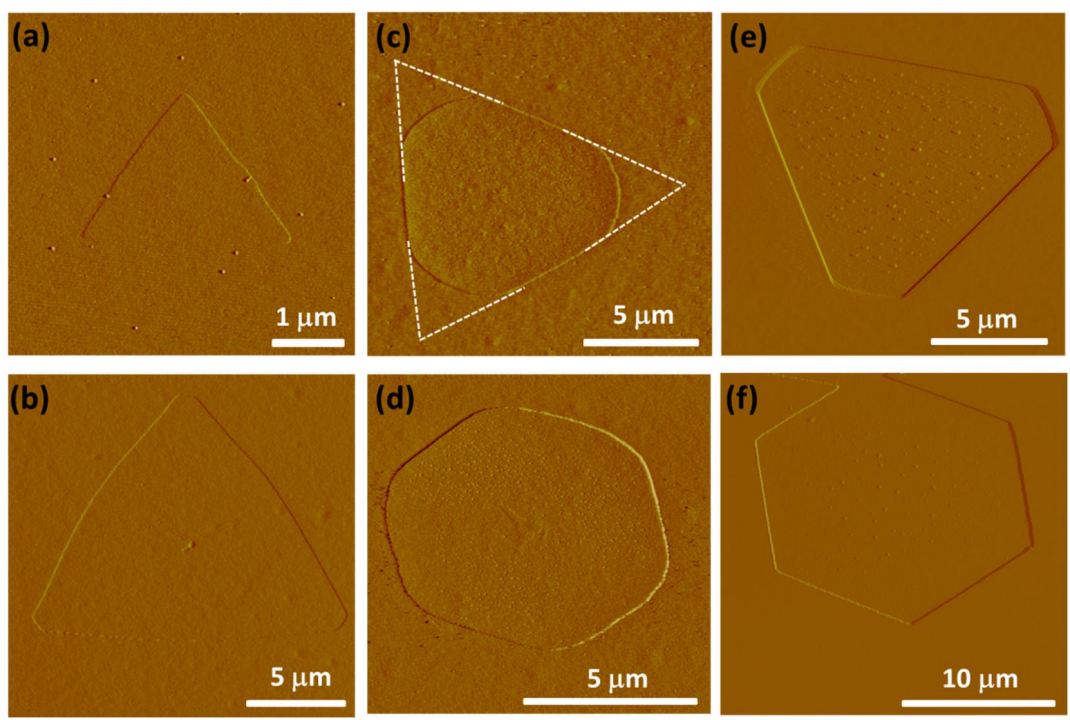

**Figure 4. Amplitude AFM images showing shape evolutions of CVD-grown WSe$_2$ flakes at different growth temperatures of (a) 900 °C, (b) 950 °C, (c) and (d) 1025 °C, (e) and (f) 1050 °C. Unusual, non-triangle shapes are gradually found as the growth temperature increases. (a), (b) are monolayer triangles with different sizes, (c), (d) are thin few layer truncated triangle and hexagon with curve edges, and (e), (f) are thick few layer triangle and hexagon with straight edges.**

Growth temperature can affect WSe$_2$ growth in several manners. For example, the sublimation speed and therefore the concentrations of WO$_{3-x}$ and Se sources, mobility and therefore diffusion rate of atoms and active species on substrates during WSe$_2$ growth, potential shift between kinetic controlled and thermodynamic controlled growth behavior during WSe$_2$ and other TMDC growth,[45] *etc*. At very low temperatures, the amount of source materials sublimated will be very few and thus the concentrations of reactants will be low. In addition, low temperature will lead to less mobile active reactants, which make them difficult to diffuse over growth substrate and difficult to add at the growing edges of 2D flakes. Instead, it is energetically preferable to grow into three dimensional structures to compensate the low mobile nature of the active species. This point is supported by our experiments that low temperature (800 °C) growth produces particle-like products. On the other hand, when the growth temperature is too high, the concentrations of reactants will be overly high, which may lead to fast growth of thick samples. This speculation is supported by our observation that high temperature growth experiments typically produce samples with greater thickness and flake sizes (Figures 3 and 4). It is also possible that monolayer WSe$_2$ is not stable at high temperatures over 1000 °C. To test this, we grew monolayer WSe$_2$ at 950 °C and annealed the samples at 1000 °C in Ar/H$_2$ (320/20 sccm) for 15 min (no WO3 or Se was introduced during annealing). After annealing, we found that monolayer WSe$_2$ undergoes severe degradation as many holes formed on the flakes and their PL has been significantly quenched. Moreover, at high temperatures, if the WSe$_2$ growth proceeds too fast, then the growth may shift from low temperature thermodynamic-controlled fashion to high temperature kinetic-controlled fashion, leading to the growth of WSe$_2$ flakes with unusual or non-equilibrium shapes (Figure 4c-4f). At sufficiently high growth temperatures, it is even possible that some low temperature prohibited reactions may be activated. Overall, these analyses can qualitatively explain the experimental trends we shown in Figures 3 and 4.

Growth of large size domains is an important topic in 2D material synthesis, including graphene, hexagonal boron nitride, and TMDCs. Typically, the nucleation density should be low and a long growth time is necessary to produce large domains of 2D materials.[47, 48] Here, we studied the effects of growth durations on the sizes and other properties of WSe$_2$ at a growth temperature of 950 °C, as this temperature is suitable to grow solely WSe$_2$ monolayers. Figure 5a-5c shows optical microscopy images of WSe$_2$ flakes grown for 1 min, 5 min, and 5 hrs, respectively. One can clearly see that the flakes grown at 1 min are quite small, with average size of only 2.5 μm. This average size increases to 4.5 μm for 5 min grown samples. The samples grow at 15 min and beyond show similar average sizes of around 12 μm, which are much larger than those flakes grow at short durations (Figure 5d). The largest flakes we found are around 20 μm for monolayer and 40 μm for few layer WSe$_2$ (grown at 1050 °C). One can see from Figure 5c that for WSe$_2$ grown for 5 hrs, the flakes maintain the triangular shapes. We also used Raman and PL to characterize the properties of WSe$_2$ flakes grown for 5 hrs (see Figure S2 in the Supporting Information) and the results show that the flakes are still monolayers, the same with the samples grown at short durations. This result suggests that at a fixed growth temperature and amount of source materials, increasing growth time will not change the layer number and shapes of WSe$_2$, instead, it will increase their lateral sizes within certain period. This point is understandable because the layer numbers and shapes of 2D flakes are mainly related to the concentrations of the source materials and growth kinetics of WSe$_2$ flakes, which are sensitive to the growth temperatures and mass of source materials, not the growth durations. Therefore, similar to graphene, increasing growth time should be an effective way to grow large size single crystalline domains of monolayer WSe$_2$.

However, there is one key difference between CVD growth of WSe$_2$ (also other TMDCs) and graphene. In WSe$_2$ growth, we find that the flakes sizes stop increasing after a relative short growth period of ~15 min, which is different with graphene growth since graphene can keep growing for as long as many hours.[47, 48] Cease of WSe$_2$ growth in short time period is related to the use of solid and non-sustainable source materials (WO$_3$ and Se powders) during WSe$_2$ synthesis. The amount of source materials will gradually decrease as the growth proceeds, leading to the slowing down and eventually ceasing of WSe$_2$ growth after certain period. This point is also reflected by the observation that there will be no Se powders left after a growth duration of 15 min. This feature is quite different from graphene growth, where the concentrations of gas phase carbon sources can be identical through the long growth period, leading to the growth of very large single domain graphene flakes up to millimeter to centimeter scale. Exploring gas phase source materials,[38] or feeding solid W and Se source materials in a continuous way, should be an effective strategy to grow large domain single crystalline $WSe_2$ and other TMDC flakes.

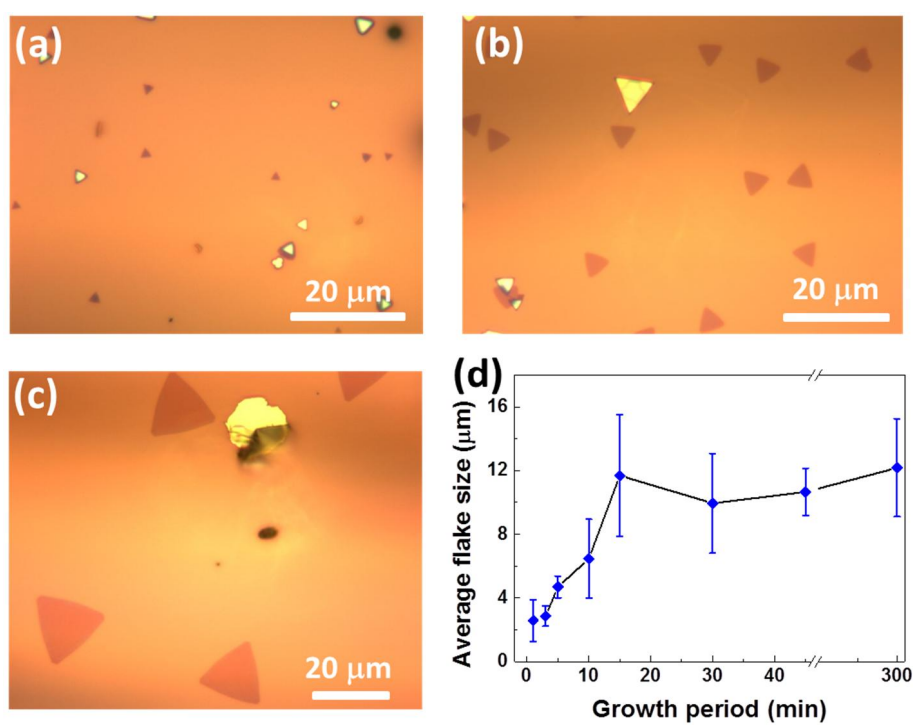

**Figure 5. Effect of growth durations on the sizes of CVD-grown monolayer $WSe_2$. Optical microscopy images of $WSe_2$ grown for (a) 1 min, (b) 5 min, and (c) 5 hrs. The growth temperatures are 950 °C for all cases. (d) Plot of average flake sizes versus growth durations of 1 min, 3 min, 5 min, 10 min, 15 min, 30 min, 60 min, and 5 hrs. The vertical error bars are standard deviations in statistical analysis.**

**Conclusions**

In summary, we report CVD growth of 2D $WSe_2$ flakes directly on $Si/SiO_2$ substrates using $WO_3$ and Se powders as source materials. We show that flake sizes and layer numbers of as-grown $WSe_2$ increase as increasing growth temperatures, leading to the growth of single domains of monolayer and few layer $WSe_2$ with sizes of ~20 μm and ~40 μm, respectively. Moreover, we observed that in addition to normal triangular $WSe_2$ flakes, some unusual shapes gradually appear, including truncated triangles and hexagons with different edge features, revealing a critical effect of temperature on the shapes and edge structures of $WSe_2$. We also find that long growth duration leads to the growth of $WSe_2$ with increased domain sizes while retaining their monolayer property and triangle shape, suggesting a practical way to grow large single crystalline $WSe_2$ by continuously feeding of precursors. FETs fabricated using such CVD-grown monolayer $WSe_2$ show a p-type behavior when using Pd as contact metal, while they exhibit an ambipolar behavior when using either Au or Ti as contact metals. These results point out the potential of using these CVD-grown monolayer $WSe_2$ materials for electronics and optoelectronics with tunable properties.

**Methods**

**CVD growth of monolayer and few layer $WSe_2$.** We used a three zone furnace for CVD growth of $WSe_2$. A schematic of our CVD set up is shown in Figure S3 in the Supporting Information. Specifically, Se powders (440 mg, 99.5%, Sigma Aldrich) were put in the first zone at upstream, and $WO_3$ powders (260 mg, 99.9%, Sigma

Aldrich) were put in the third zone. The distance between the two sources was tuned and optimized at a long distance of 55 cm. We speculate that via transport through such a long distance, the concentrations of Se will be relatively uniform along length scale of a growth substrate (~2 cm), which would be benefit to the growth of uniform $WSe_2$ on substrate. A 2 inch quartz reaction tube was first flushed with 2000 sccm of Ar for 10 min and then the furnace was ramped to the designed temperature at a ramp rate of 50 °C/min for growth. The temperatures of Se and $WO_3$ were controlled by their locations in the three zone furnace over wide ranges. The results presented in Figure 1 were grown at conditions where the temperature of $WO_3$ was 950 °C, and Se was 540 °C. The growth substrates were silicon wafers with 300 nm $SiO_2$, which were put right on top of $WO_3$ powders and were facing down. During growth process, the flow rate of $Ar/H_2$ was tuned and optimized at 320/20 sccm, and the growth was at ambient pressure. After reaction, the furnace was naturally cooling down to below 200 °C under 320/20 sccm of Ar/H2, and the samples were taking out for characterization. We found that there was no $WSe_2$ growth if $H_2$ was not introduced in the CVD process. This observation is consistent with some recent reports.[35, 36] The use of reductive species like $H_2$ or sulfur may help sublimation of $WO_{3-x}$ to increase its concentration in vapor phase,[35, 36] thus promoting $WSe_2$ growth. We also tried low pressure growth (2-100 Torr), but these low pressure experiments usually lead to very few deposits on substrates, presumably due to reduced concentrations of reactants at low pressure environment.

**Characterization.** The samples were characterized using optical microscopy, AFM (Dimensional 3100, Digital Instruments, tapping mode), and Raman and PL spectroscopy (Renishaw Raman with a 532 nm laser). The laser spot size was around 1 μm during Raman and PL measurements.

**Device fabrication and measurements.** Back gated $WSe_2$ FETs were directly fabricated on $Si/SiO_2$ substrates where $WSe_2$ monolayers were grown on. The devices were fabricated using e-beam lithography and different contact metals were studied, included Pd/Ti (50 nm/0.5 or 1 nm), Au/Ti (50 nm/0.5 or 1 nm), and Ti/Au (5 nm/50 nm). The device measurements were conducted in ambient condition using an Agilent 4156B Semiconductor Parameter Analyzer.

*Conflict of Interest:* The authors declare no competing financial interest.

*Acknowledgement.* We thank Han Wang for helpful discussions. We would like to acknowledge the collaboration of this research with King Abdul-Aziz City for Science and Technology (KACST) via The Center of Excellence for Nano-technologies (CEGN). We also acknowledge support from the Office of Naval Research (ONR) and the Air Force Office of Scientific Research (AFOSR).

*Supporting Information Available:* Additional Raman results. This material is available free of charge via the Internet at http://pubs.acs.org.

**TOC Figure**

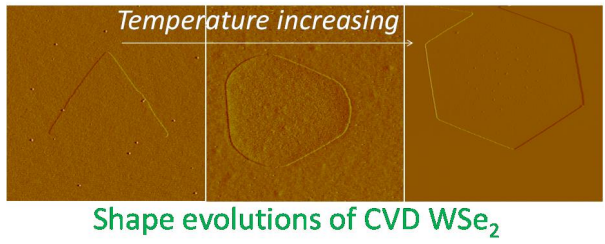

Shape evolutions of CVD WSe$_2$

**Supporting Information:** *ACS Nano*

**Chemical Vapor Deposition Growth of Monolayer WSe$_2$ with Tunable Device Characteristics and Growth Mechanism Study**

Bilu Liu[+]*, Mohammad Fathi[+], Liang Chen, Ahmad Abbas, Yuqiang Ma, Chongwu Zhou*

Ming Hsieh Department of Electrical Engineering, University of Southern California, Los Angeles, California 90089, USA

[+]Equal contribution

Address correspondence to: chongwuz@usc.edu, biluliu@usc.edu

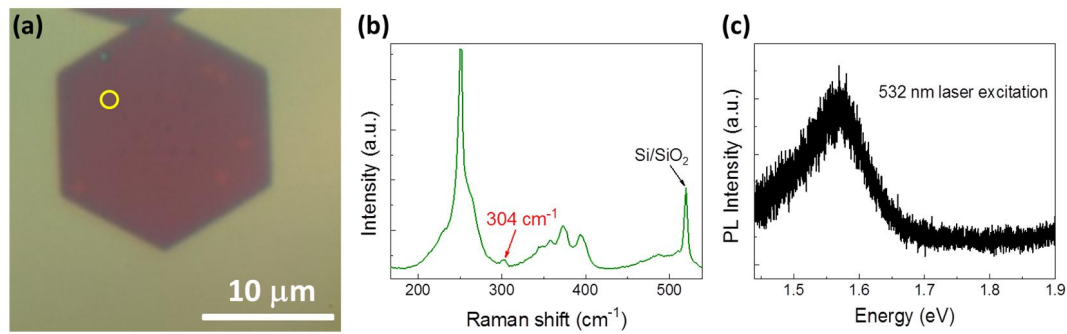

**Figure S1.** Characterization of a few layer hexagonal WSe$_2$ flake grown at 1050 °C. (a) An optical microscopy image of the same flake shown in Figure 4f. (b), (c) Raman and PL spectra taken from the hexagonal flake (yellow circle position in the image a). The existence of the $B_{2g}^1$ peak at 304 cm$^{-1}$ in Raman spectrum[1] and very weak PL intensity reveal that this flake is a few layer WSe$_2$.

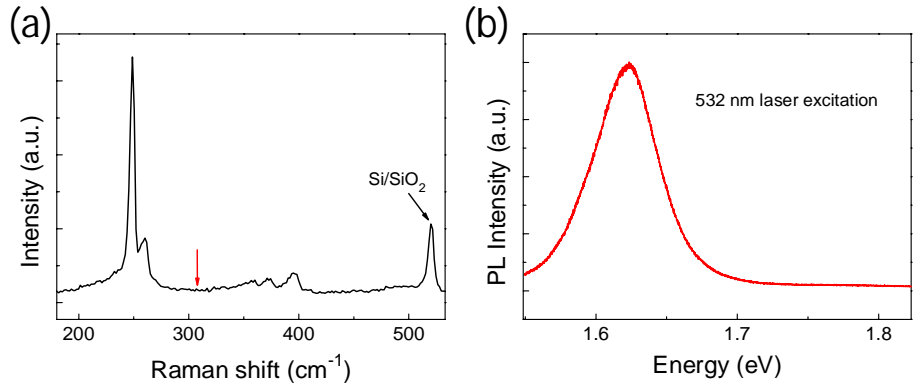

**Figure S2.** Typical Raman spectrum (a) and PL spectrum (b) of WSe$_2$ flakes grown at 950 °C for a long growth duration of 5 hrs. The absence of the B$_{2g}^1$ Raman peak at ~304 cm$^{-1}$ and strong PL intensity reveal that the samples are monolayer WSe$_2$ flakes.

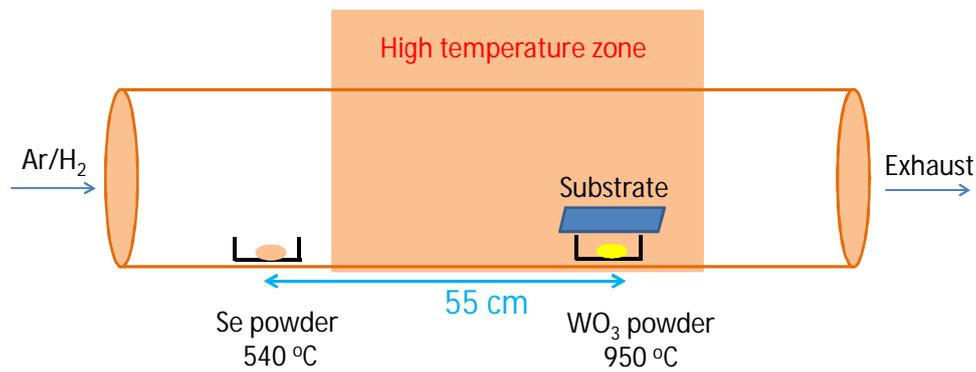

**Figure S3.** A schematic of the three-zone CVD setup used for the growth of WSe$_2$ flakes in this study.

*References.*